\documentclass[a4paper]{jpconf}

\usepackage{amssymb}

\def\ba{\begin{eqnarray}}
\def\ea{\end{eqnarray}}

\def\sp{\kern +3pt}

\usepackage{graphicx}
\begin{document}
\title{A covariant model for the negative parity resonances
of the nucleon}

\author{G.~Ramalho}

\address{International Institute of Physics, Federal 
University of Rio Grande do Norte, Av.~Odilon Gomes de Lima 1722, 
Capim Macio, Natal-RN 59078-400, Brazil}

\ead{gilberto.ramalho2013@gmail.com}

\begin{abstract}
We present a model for the $\gamma^\ast N \to N^\ast$ 
helicity amplitudes,
where $N$ is the nucleon and $N^\ast$ is a 
negative parity nucleon excitation, 
member of the $SU(6)$-multiplet $[70,1^-]$. 
The model combines the results from the  
single quark transition model for the helicity 
amplitudes with the results of the covariant spectator quark 
model for the $\gamma^\ast N \to N^\ast(1535)$ 
and $\gamma^\ast N \to N^\ast(1520)$ transitions.
With the knowledge of the amplitudes $A_{1/2}$ and $A_{3/2}$ 
for those transitions we calculate three independent coefficients 
defined by the single quark transition model
and make predictions for the helicity amplitudes  
associated with the 
$\gamma^\ast N \to N^\ast(1650)$,  
$\gamma^\ast N \to N^\ast(1700)$,
$\gamma^\ast N \to \Delta(1620)$,
and $\gamma^\ast N \to \Delta(1700)$ transitions.
In order to facilitate the comparison with 
future experimental data at high $Q^2$,
we provide also simple parametrizations for the amplitudes,  
compatible with the expected falloff at high $Q^2$.
%\vspace{1.cm}
\end{abstract}

\section{Introduction}

\begin{figure}%[c]
%\vspace{1.cm}
%\vspace{.6cm}
%\centerline{
%\mbox{
%\includegraphics[width=9.6cm]{SigmaW2.eps} %}}
\includegraphics[width=9.6cm]{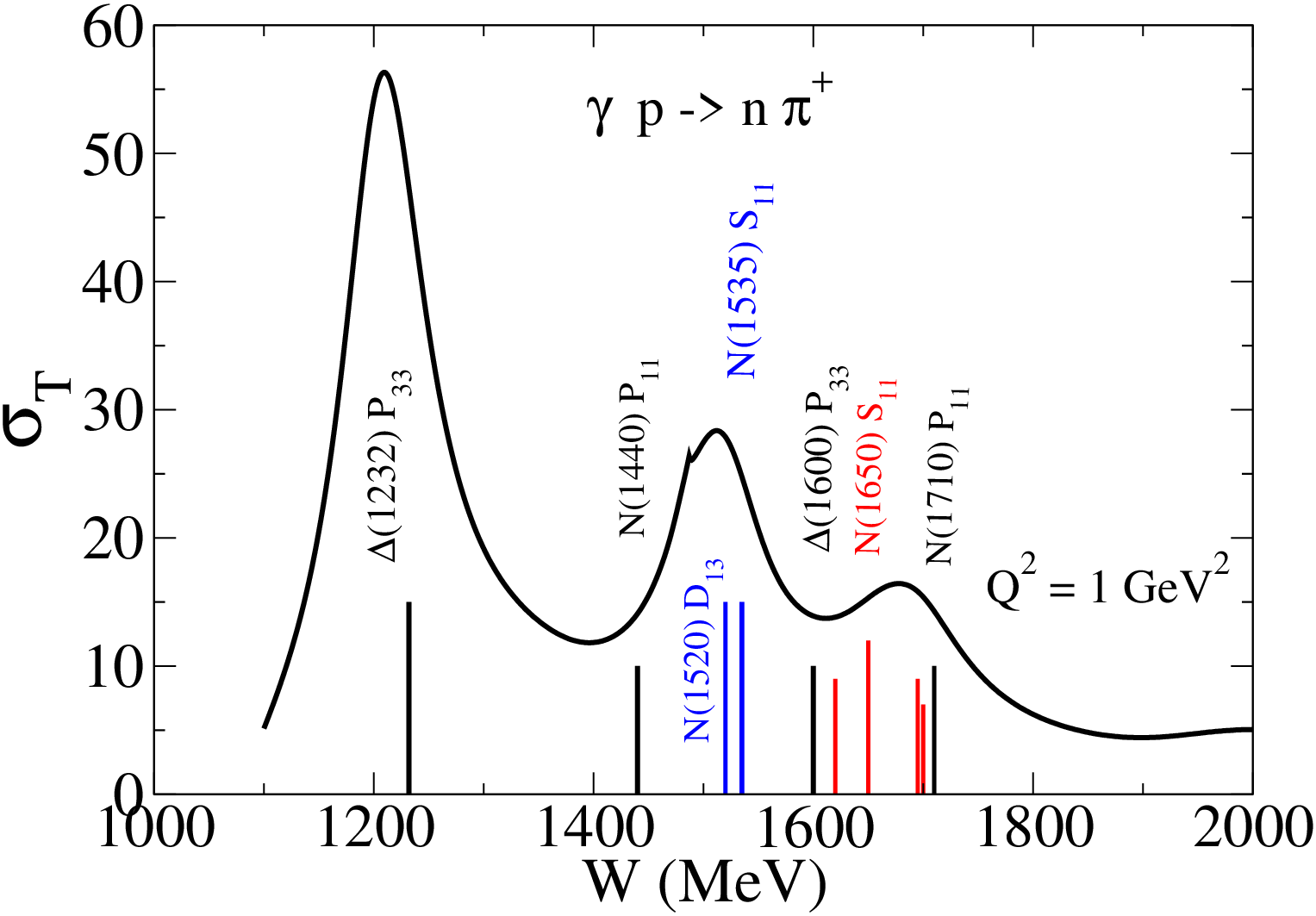} %}}
%\vspace{-1cm}
\hspace{.3cm}
\begin{minipage}[b]{12pc}\caption{\label{figSigmaW}
Representation of the $\gamma p \to n \pi^+$ 
cross section.
The graph define the 3 resonance regions.
The vertical lines 
represent resonant states 
described by the covariant spectator quark model.
At red we indicate the states studded 
in this work.
At blue are the states used as input.}
\end{minipage}
\end{figure}

One of the challenges in the modern physics 
is the description of the internal structure 
of the baryons and mesons.
The electromagnetic structure 
of the nucleon $N$ and the nucleon resonances $N^\ast$ can be accessed 
through the $\gamma^\ast N \to N^\ast$ reactions,
which depend  of the (photon) 
momentum transfer squared $Q^2$~\cite{NSTAR,Aznauryan12,CLAS,MAID}. 
The data associated with those transitions 
are represented in terms of   
helicity amplitudes and have been  
collected in the recent years at Jefferson Lab, 
with increasing $Q^2$~\cite{NSTAR}.
The new data demands the development of theoretical models 
based on the underlying structure 
of quarks and quark-antiquark states (mesons)~\cite{NSTAR,Aznauryan12}.
Those models may be used to guide future 
experiments as the ones planned for the Jlab--12 GeV upgrade, particularly 
for resonances in the second and third resonance region
[energy $W =1.4$--$1.8$ GeV]
(see figure~\ref{figSigmaW})~\cite{NSTAR}.
In that region there are several resonances $N^\ast$ 
from the multiplet $[70,1^-]$ of $SU(6)\otimes O(3)$,
characterized by a negative parity~\cite{Aznauryan12,Capstick00,Burkert03}.
According with the single quark transition model (SQTM),
when the electromagnetic interaction 
is the result of the photon coupling with just one quark,
the helicity amplitudes of the   $[70,1^-]$ members
depend only on three independent functions of $Q^2$:
$A,B$ and $C$~\cite{Burkert03,SQTM}.
In this work we use the 
covariant spectator quark model~\cite{NSTAR,SQTM,Nucleon} 
developed for the $\gamma^\ast N \to N^\ast (1520)$
and $\gamma^\ast N \to N^\ast (1535)$ transitions,
also members of $[70,1^-]$, to calculate those functions~\cite{S11,D13}.
Since the covariant spectator quark model 
breaks the $SU(2)$-flavor symmetry, 
we restrict our study to reactions with proton targets
(average on the SQTM coefficients)~\cite{SQTM}.
Later on, with the knowledge of the functions $A,B$, and $C$
we predict the helicity amplitudes 
for transitions associated 
with the remaining members of the multiplet $[70,1^-]$~\cite{SQTM}.

%\vspace{-.1cm}

\section{Covariant Spectator Quark Model}

The covariant spectator quark model is based 
on the formalism 
of the covariant spectator theory~\cite{Gross}.
In the covariant spectator quark model,
baryons are treated as
three-quark systems. 
The baryon wave functions are derived from 
the quark states according with the 
$SU(6) \otimes O(3)$ symmetry group.
A quark is off-mass-shell, and  free
to interact with the photon fields,
and other two quarks are on-mass-shell~\cite{Nucleon,Nucleon2,OctetFF,Omega}.
Integrating over the quark-pair degrees
of freedom we reduce the baryon to a quark-diquark system,
where the diquark can be represented as
an on-mass-shell spectator particle with an effective
mass $m_D$~\cite{Nucleon,S11,D13,Nucleon2,Omega}.

The electromagnetic interaction with the baryons
is described by the photon coupling with
the constituent quarks in the relativistic impulse approximation.
The quark electromagnetic structure is
represented in terms of the quark form
factors parametrized by a vector meson dominance
mechanism~\cite{Nucleon,Omega,Lattice}.
The parametrization of the quark current
was calibrated in the studies of the nucleon form factors data~\cite{Nucleon},
by the lattice QCD data for the decuplet baryon~\cite{Omega},
and encodes effectively the gluon
and quark-antiquark substructure of the constituent quarks.
The quark current has the general form~\cite{Nucleon,Omega}
\ba
j_q^\mu(Q^2) = j_1(Q^2) \gamma^\mu
+ j_2(Q^2) \frac{i \sigma^{\mu \nu} q_\nu}{2M},
\label{eqJq}
\ea
where $M$ is the nucleon mass and
$j_i$ $(i=1,2)$ are the Dirac and Pauli quark form factors.
In the $SU(2)$-flavor sector
the functions $j_i$ can also be decomposed 
into the isoscalar ($f_{i+}$) and the isovector ($f_{i-}$) 
components: 
$j_i = \frac{1}{6} f_{i+} + \frac{1}{2} f_{i-} \tau_3$,
where $\tau_3$ acts on the isospin states of baryons
(nucleon or resonance).
The details can be found in~\cite{Nucleon,OctetFF,Omega}.

When the nucleon wave function ($\Psi_N$) 
and the resonance wave function ($\Psi_R$)
are both expressed in terms of
the single quark and quark-pair states,
the transition current in impulse approximation as
can be  written~\cite{Nucleon,Nucleon2,Omega}
\ba
J^\mu=
3 \sum_{\Gamma} 
\int_k \bar \Psi_R (P_+,k) j_q^\mu \Psi_N(P_-,k),
\label{eqJmu}
\ea  
where $P_-,P_+$, and $k$ are  the
nucleon, the resonance, and the diquark momenta respectively.
In the previous equation 
the index $\Gamma$ labels
the possible states of the intermediate diquark polarizations,
the factor 3 takes account of the contributions from
the other quark pairs by the symmetry, and the integration
symbol represents the covariant integration over the 
diquark on-mass-shell momentum.
In the study of inelastic reactions 
 we replace $\gamma^\mu \to \gamma^\mu - \frac{\not q q^\mu}{q^2}$
in equation~(\ref{eqJq}). 
This procedure ensures the conservation 
of the transition current and 
it is equivalent to the use of 
the Landau prescription~\cite{SQTM,S11,D13}.

Using equation~(\ref{eqJmu}), we can express 
the transition current 
in terms of the 
quark electromagnetic form factor $f_{i\pm}$ ($i=1,2$)
and the radial wave functions 
$\psi_N$ and $\psi_R$~\cite{Nucleon,S11,D13}.
The radial wave functions are scalar functions that 
depend on the  baryon ($P$) and diquark ($k$) momenta
and parametrize the momentum distributions 
of the quark-diquark systems.
From the transition current we can extract 
the form factors and the helicity transition amplitudes,
defined in the rest frame of the resonance (final state), 
for the reaction under study~\cite{NSTAR,Aznauryan12,S11,D13}.

There are however some processes such as
the meson exchanged between the different quarks
inside the baryon, which cannot be reduced
to simple diagrams with quark dressing.  
Those processes are regarded  
as arising from a meson exchanged between
the different quarks inside
the baryon
and can be classified as meson cloud corrections 
to the hadronic reactions~\cite{D13,OctetFF,Octet2Decuplet}.

The covariant spectator quark model was used already 
in the study of several nucleon excitations 
including isospin 1/2 systems 
$N(1410),N(1520),N(1535),N(1710)$~\cite{S11,D13,Roper}
and the isospin 3/2 systems ~\cite{LatticeD,Delta2,Delta1600}.
The model generalized to the $SU(3)$-flavor sector 
was also used to study the octet and decuplet baryons 
as well as transitions between baryons 
with strange quarks~\cite{Octet2Decuplet,Strange}.
In figure~\ref{figSigmaW} the position 
of the nucleon excitations are represented and compared 
with the bumps of the cross sections.
Based on the parametrization of the quark current 
(\ref{eqJq}) in term of the vector meson dominance mechanism, 
the model was  extended to the lattice QCD regime 
(heavy pions and no meson cloud)~\cite{Lattice,LatticeD},
to the nuclear medium~\cite{OctetFF} 
and to the timelike regime~\cite{Timelike}.
The model was also used to study 
the nucleon deep inelastic scattering~\cite{Nucleon,NucleonDIS}
and the axial structure of the octet baryon~\cite{OctetAxial}.

%\vspace{-.3cm}
\vspace{-.2cm}

\section{Results for $N(1535)$ and $N(1520)$}

For the study of the states $N(1535)$ ($\frac{1}{2}^-$) 
and  $N(1520)$ ($\frac{3}{2}^-$) it is necessary 
to specify the shape of the radial wave 
function of the nucleon and resonant states.
The radial wave function can be represented 
using the dimensionless variable 
$\chi= \frac{(P-k)^2- (M_B-m_D)^2}{M_B m_D}$,
where $M_B$ is the mass of the baryon $B$.
We choose in particular
\ba
\psi_N(P,k)= \frac{N_0}{m_D ( \beta_2 + \chi)(\beta_1 + \chi)},
\hspace{1cm}
\psi_{R}(P,k)=
\frac{N_1}{m_D(\beta_2 + \chi)}
\left[
\frac{1}{\beta_1 + \chi} - \frac{\lambda_{R}}{\beta_j + \chi}
\right],
\label{eqPsiRadial}
\ea
where $N_0,N_1$ are normalization constants,
$\beta_1, \beta_2$ and $\beta_j$ 
are momentum range parameters in units $M_B m_D$.
We use $\beta_j=\beta_3$ for $N(1535)$ 
and $\beta_j=\beta_4$ for $N(1520)$.
Those parameters are fixed by a fit to 
the the large $Q^2$ data~\cite{SQTM,D13}. 
The coefficient $\lambda_R$ is determined by 
an orthogonality condition between 
the nucleon and the state $R$.
In the following we will use also 
the spectroscopic notation 
to represent the states $N(1535)$ (or $S11$) and $N(1535)$ (or $D13$).
In addition we use $M_S$ and $M_D$
to represent the $S11$ and $D13$ masses, respectively.

\subsection{State $N(1535)$}

From the study of~\cite{SQTM,S11}
we conclude that we can write 
the amplitude $A_{1/2}$ for the $N(1535)$ state
in terms of the Dirac transition 
form factor ($F_1^\ast$).
The final result is then
\ba
A_{1/2} = -
\frac{\sqrt{2}}{3} 
F_S\left(
f_{1+} + 2 f_{1-} \tau_3 \right) {\cal I}_{S11} \cos \theta_S,
\hspace{1cm}
{\cal I}_{S11} (Q^2)=  \int_k  
\frac{k_z}{|{\bf k}|}
\psi_{S11}(P_+,k) \psi_N(P_-,k), 
\label{eqA12-S11}
\ea
where
$F_S= 2 e \sqrt{\frac{(M_S + M)^2+Q^2}{8M(M_S^2-M^2)}}$
and ${\cal I}_{S11}$ 
is a covariant integral calculated 
on the $S11$ rest frame.
%In the $S11$ rest frame one has $P_+= (M_S,0,0,0)$ 
%and $P_-=(\sqrt{M^2 + |{\bf q}|^2}, 0,0 ,-|{\bf q}|)$,
%where $|{\bf q}|$ is the magnitude of the photon three-momentum.

The result (\ref{eqA12-S11}) is valid for large $Q^2$
since only valence quark effects are considered
and it is the consequence 
of the observation that the Pauli transition form factor ($F_2^\ast$)
vanishes for $Q^2 \gtrsim 1.5$ GeV$^2$,
which is interpreted as the consequence of the cancellation between 
valence quark and meson cloud effects~\cite{S11}.
In this work we updated the model from~\cite{S11}
using the radial wave function $\psi_R$ given by  
equation (\ref{eqPsiRadial}), in order to ensure 
the exact orthogonality between nucleon and $N(1535)$ states.
In the process we introduce a new parameter 
($\beta_3$) that is adjusted by the large $Q^2$ data 
(no meson cloud) \cite{SQTM}.
%

%\clearpage
%\input{sec32_v2}
%\clearpage

%\vspace{-.3cm}
%\vspace{-.15cm}
\vspace{-.2cm}

\subsection{State $N(1520)$}

The model for  from~\cite{D13}
can be used to calculate the electromagnetic transition 
form factors for the $\gamma^\ast N \to N(1520)$ transition,
including the magnetic dipole $G_M$ 
and the electric quadrupole $G_E$ form factors,
based in the effects of the valence quarks.
One obtain then $G_E = -G_M$, where 
\ba
G_M= 
{\cal R}(f_{1+} + 2 f_{1-} \tau_3)  {\cal I}_{D13}
+ \frac{M_D + M}{2M}  
(f_{2+} + 2 f_{2-} \tau_3) {\cal I}_{D13}.
\ea
In the equations
${\cal R}= \frac{1}{3\sqrt{3}} \frac{M}{M_D-M} 
\sqrt{\frac{(M_D -M)^2 + Q^2}{(M_D + M)^2 + Q^2}}$, and
${\cal I}_{D13} (Q^2)=  \int_k  
\frac{k_z}{|{\bf k}|}
\psi_{D13}(P_+,k) \psi_N(P_-,k)$
is the new invariant integral defined  at 
the resonance rest frame.

\begin{figure}
%\vspace{-.2cm}
%\centerline{
%\mbox{
\includegraphics[width=2.0in]{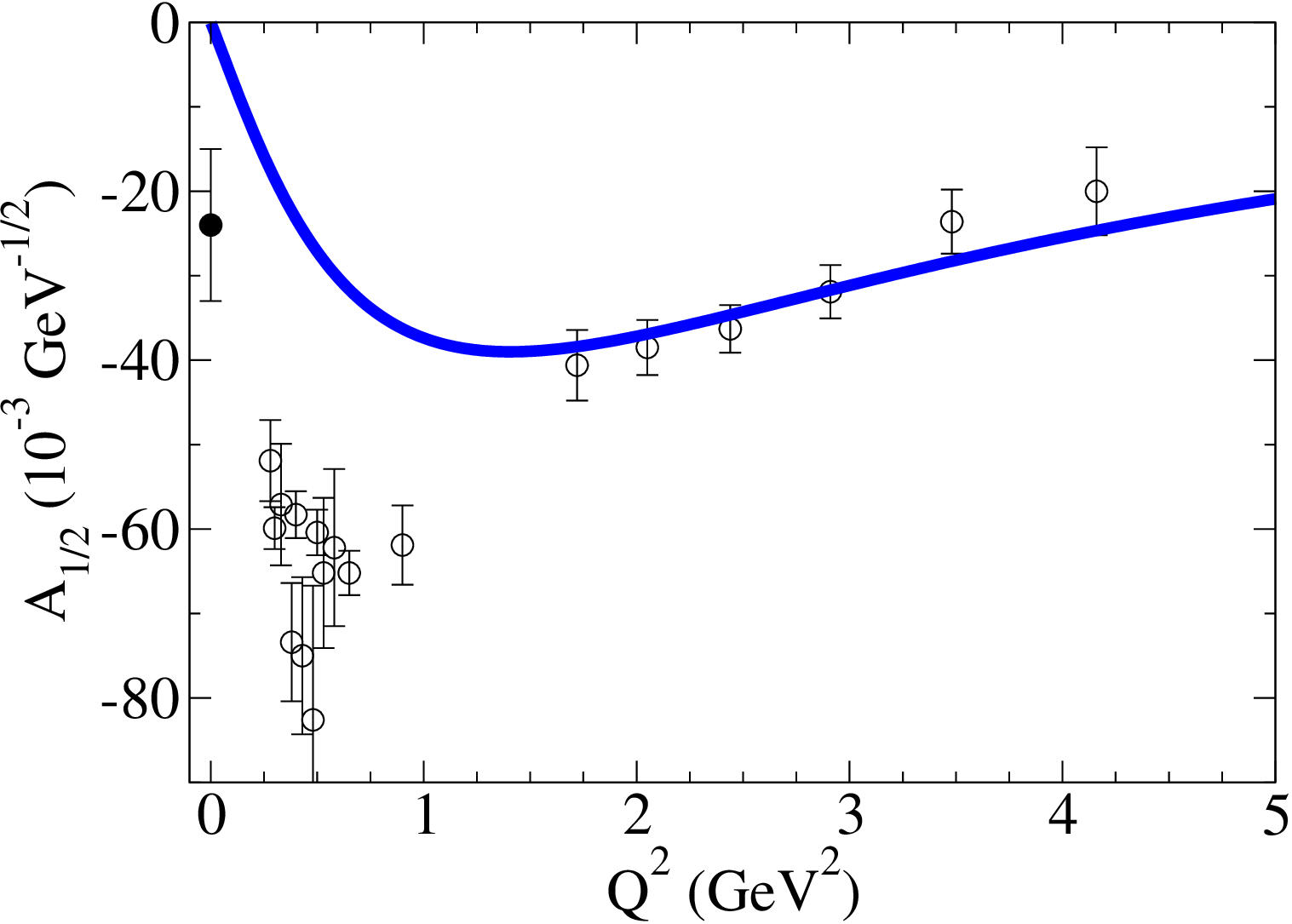}  \hspace{.1cm}
\includegraphics[width=2.0in]{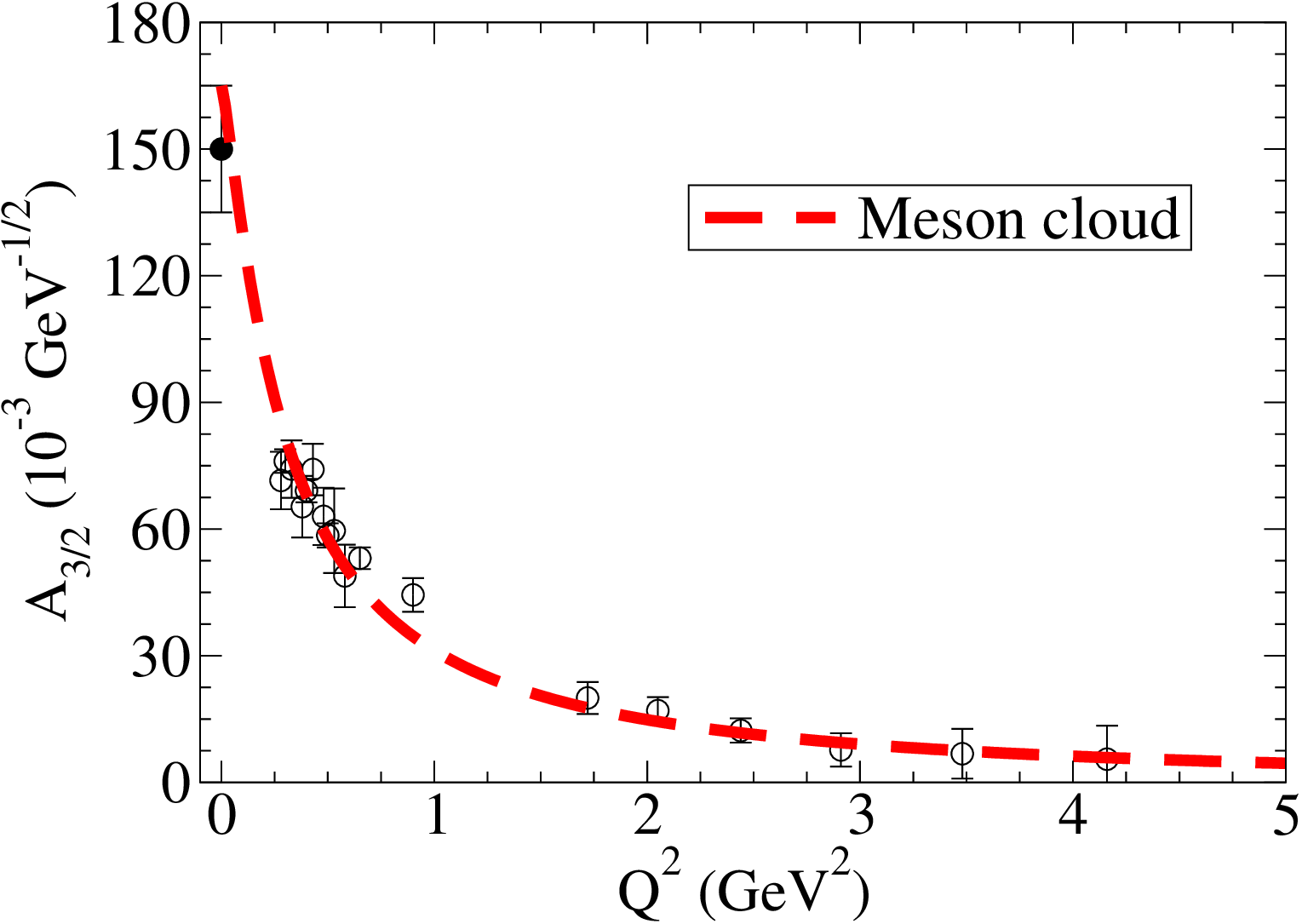} %}
\hspace{.3cm}
\begin{minipage}[b]{10pc}\caption{\label{figN1520}
Results for the resonance $N(1520)$.
The amplitude $A_{3/2}$ the result 
of meson cloud effects. 
was no contributions from valence quarks.
Data from CLAS~\cite{CLAS}.}
\end{minipage}
\end{figure}

The result $G_M + G_E=0$ is interesting, 
since it is consistent with the expected QCD behavior 
for large $Q^2$, but it is inconsistent with the data at low $Q^2$,
that shows a significant magnitude for the  
amplitude $A_{3/2} \propto (G_M + G_E)$,
near $Q^2=0$~\cite{Aznauryan12,SQTM,D13}.
In general, quark models 
predict small contributions for $A_{3/2}$ 
(20--40\%)~\cite{SQTM,D13}.
There are however indications that 
the effects of the meson cloud contribution
dominate the amplitude, 
as supported by the calculation 
from  EBAC at Jefferson Lab~\cite{EBAC}. 
Based on that information we represent 
the helicity amplitudes as
\ba
A_{1/2}= {F_D} \, G_M + \frac{1}{4}F_D \, G_4^\pi, 
\hspace{1cm}
A_{3/2}= \frac{\sqrt{3}}{4} F_D \, G_4^\pi, 
\label{eqAmps-D13}
\ea
where
$F_D=  \frac{e}{M} \sqrt{\frac{M_D-M}{M_D+M}} 
\sqrt{\frac{(M_D+M)^2 + Q^2}{2M}}$.
In the equations~(\ref{eqAmps-D13}), 
$G_4^\pi$ is a function that is not determined by 
the covariant spectator quark model 
(that predicts $G_4^\pi \equiv 0$) 
and parametrize the amplitude $A_{3/2}$
assuming the dominance of the pion/meson cloud effects.
The function $G_4^\pi$ is fitted to the data 
with a model inspired on the pion and meson 
cloud contributions for the $\gamma^\ast N \to \Delta$
transition~\cite{SQTM,D13,Timelike}.
The results from the amplitudes 
$A_{1/2}$ (valence quark) and 
$A_{3/2}$ (meson cloud) are presented on figure~\ref{figN1520}.

%\vspace{-.2cm}

\vspace{-.15cm}

\section{Single Quark Transition Model}

The combination of the 
wave functions of a baryon (three-quark system)
given by  $SU(6)\otimes O(3)$ group 
and the description of 
electromagnetic interaction in impulse approximation 
leads to the so-called single quark transition model %(SQTM) 
\cite{Burkert03,SQTM-refs}.
In this context {\it single} means 
that only one quark couples with the photon. 
In these conditions the SQTM can be used to 
parametrize the transition current 
between two multiplets,
in an operational form 
that includes only four independent terms,
with coefficients exclusively dependent of $Q^2$.

In particular, the SQTM can be used to parametrize 
the $\gamma^\ast N \to N^\ast$  transitions,
where $N^\ast$ is a $N$ (isospin 1/2) or a $\Delta$ 
(isospin 3/2) state from the
$[70,1^-]$ multiplet, in terms 
on three independent functions of $Q^2$:
$A,B$, and $C$ \cite{Burkert03,SQTM-refs}.
The relations between the functions $A,B$, and $C$
and the amplitudes are presented in the Table~\ref{tableAmp}.
Using the results for the 
$\gamma^\ast N \to N (1535)$ 
and $\gamma^\ast N \to N(1520)$ amplitudes, 
respectively $A_{1/2}^{S11}$, $A_{1/2}^{D13}$, and $A_{3/2}^{D13}$
in the spectroscopic notation,
we can write
\ba
& &
A= 2 \frac{A_{1/2}^{S11}}{\cos \theta_S} 
+ \sqrt{2} A_{1/2}^{D13} +
\sqrt{6} A_{3/2}^{D13}, \hspace{1.5cm} 
%\label{eqA}
%\\& & 
B=   2 \frac{A_{1/2}^{S11}}{\cos \theta_S} 
- 2 \sqrt{2} A_{1/2}^{D13}, 
\nonumber \\ 
%\label{eqB} \\
& &
C= - 2 \frac{A_{1/2}^{S11}}{\cos \theta_S} 
- \sqrt{2} A_{1/2}^{D13} +
\sqrt{6} A_{3/2}^{D13}. 
\label{eqABC}
\ea

\begin{table}[t]
\begin{tabular}{c c c }
\hline
\hline
State & Amplitude &     \\
\hline
$N(1535)$ & $A_{1/2}$ & $\frac{1}{6}(A+B-C) \cos  \theta_S$ \\ [.3cm]
% & & \\
$N(1520)$
& $A_{1/2}$ & $\frac{1}{6\sqrt{2}}(A-2B-C)\cos \theta_D$   \\
& $A_{3/2}$ & $\frac{1}{2\sqrt{6}}(A+ C) \cos \theta_D$   \\ [.3cm]
% & & \\
$N(1650)$ & $A_{1/2}$ & $\frac{1}{6}(A+B-C) \sin  \theta_S$ \\ [.3cm]
$\Delta(1620)$ & $A_{1/2}$ & $\frac{1}{18}(3A-B+C) $ \\ [.3cm]
$N(1700)$ & $A_{1/2}$ &  $\frac{1}{6\sqrt{2}}(A-2B-C)\sin \theta_D$  \\
& $A_{3/2}$ & $\frac{1}{2\sqrt{6}}(A+ C) \sin \theta_D$   \\ [.3cm]
$\Delta(1700)$ & $A_{1/2}$ &  
$\frac{1}{18 \sqrt{2}}(3A+2B+C)$  \\ 
& $A_{3/2}$ & $\frac{1}{6\sqrt{6}}(3A-C) $   \\ [.1cm]
\hline
\hline
\end{tabular}
\hspace{.3cm}
\begin{minipage}[b]{14pc}\caption{\label{tableAmp}
Amplitudes $A_{1/2}$ and $A_{3/2}$ estimated by SQTM 
for the proton targets ($N=p$)~\cite{Burkert03,SQTM}.
The angle $\theta_S$ is the mixing angle associated 
with the %$S_{11}$ 
$N \frac{1}{2}^-$ states ($\theta_S =  31^\circ$).
The angle $\theta_D$ is the mixing angle associated 
with the %$D_{13}$ 
$N \frac{3}{2}^-$ 
states ($\theta_S = 6^\circ$).}
\end{minipage}
\end{table}

\begin{figure}
%\vspace{-.1cm}
%\vspace{.3cm}
\vspace{.1cm}
\centerline{
\mbox{%\hspace{7cm}
%\hspace{7.5cm}
\includegraphics[width=2.0in]{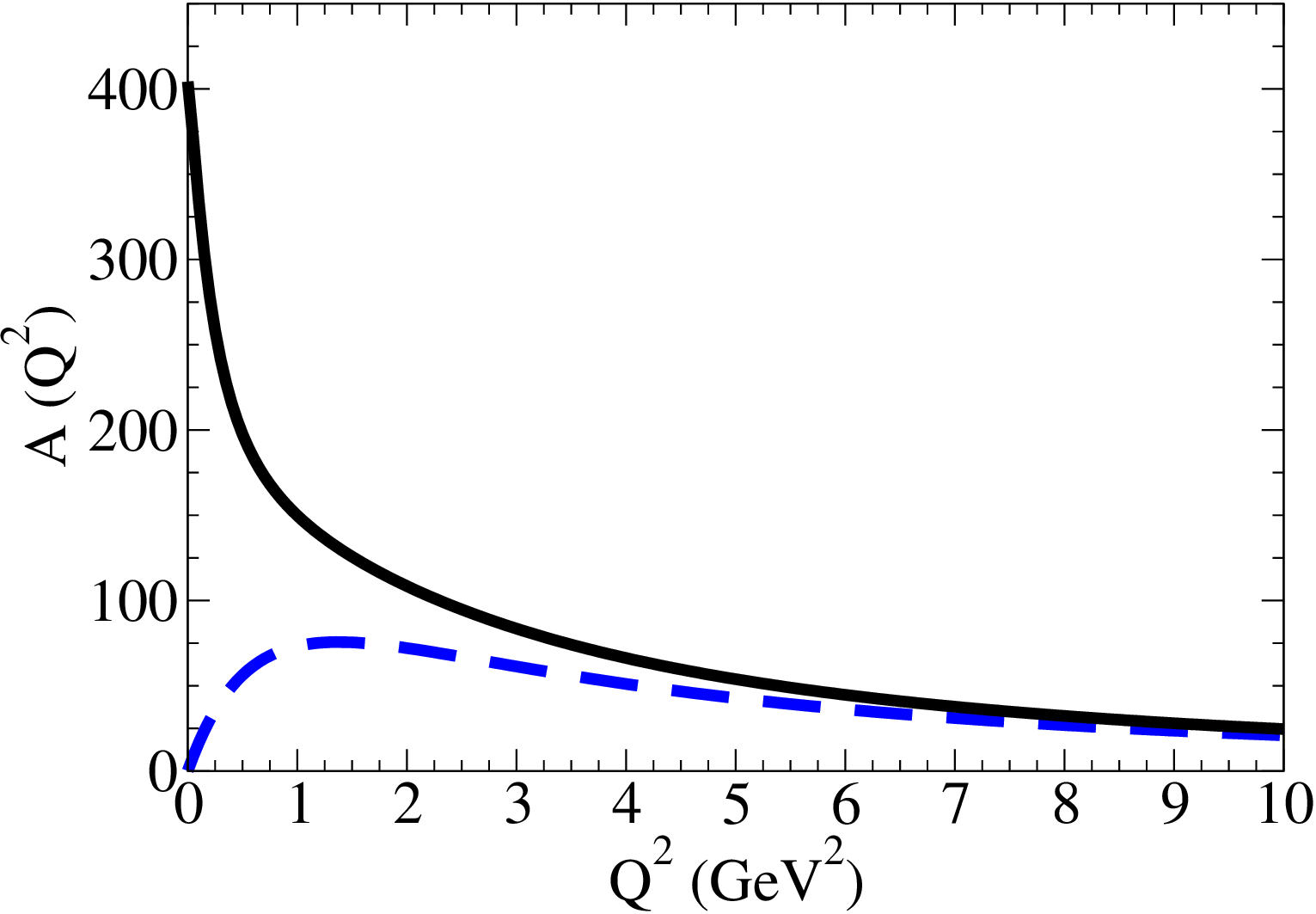}  \hspace{.1cm}
\includegraphics[width=2.0in]{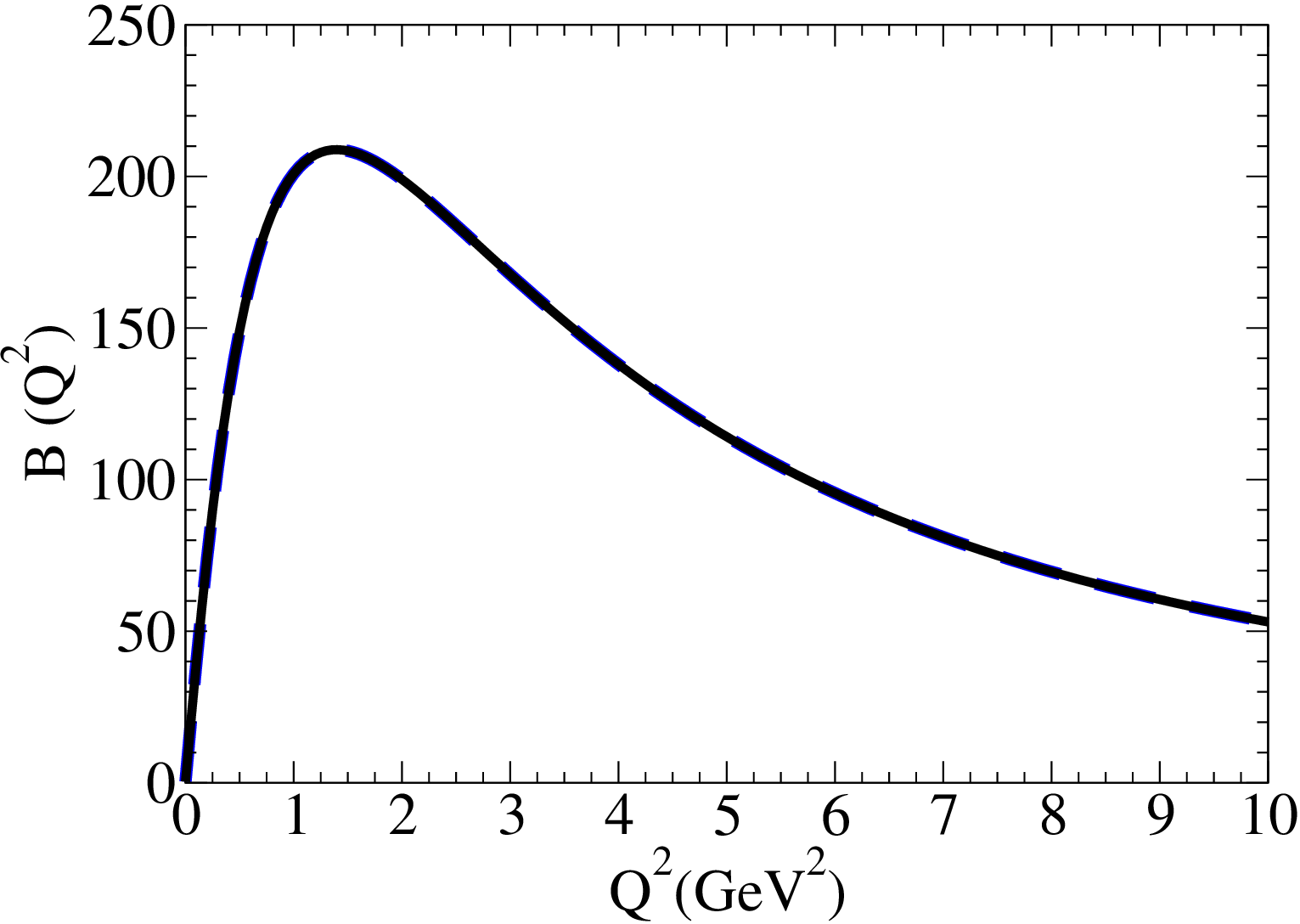}  \hspace{.1cm}
\includegraphics[width=2.0in]{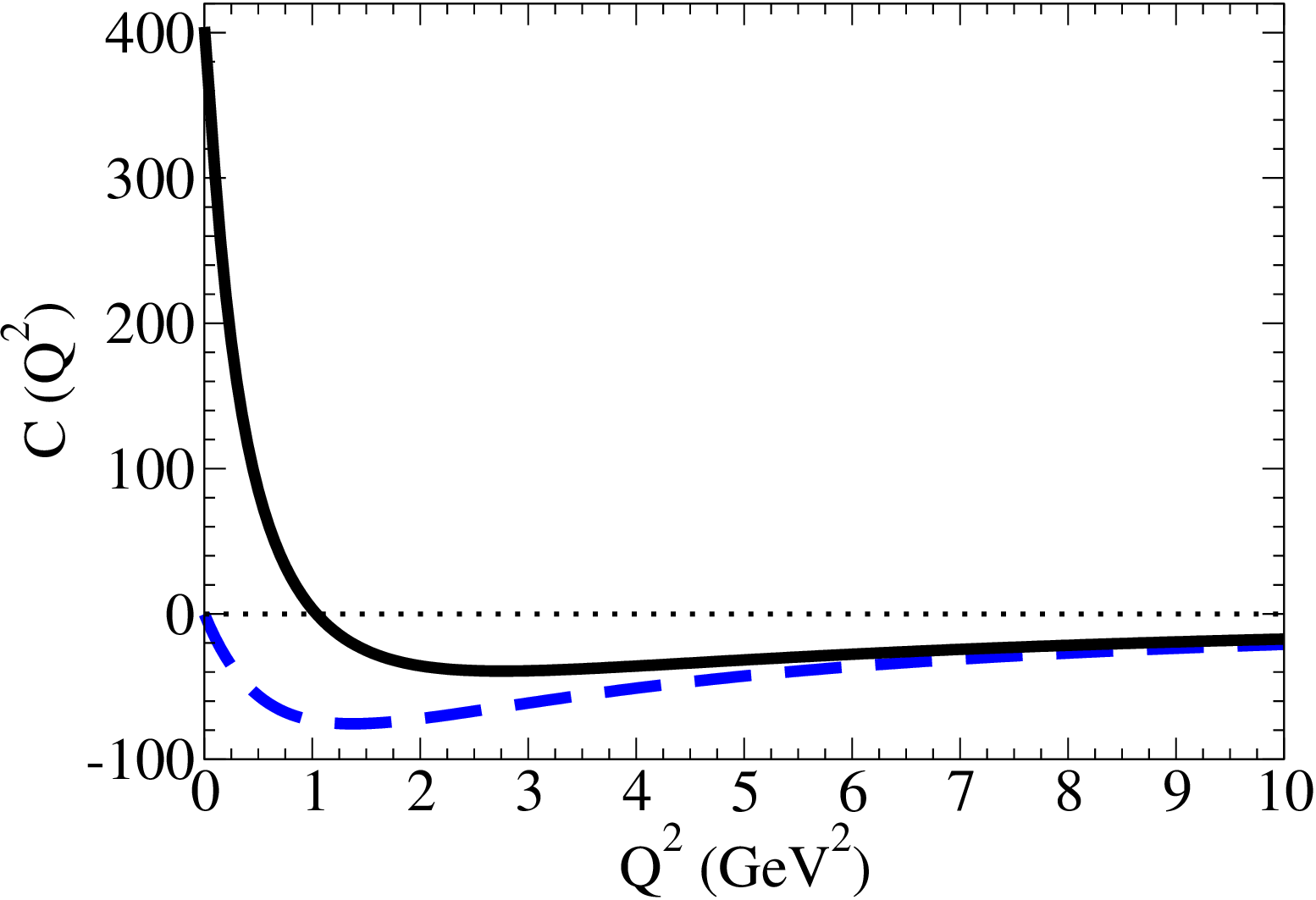}
}}
\caption{\label{figABC}
Results for the coefficients $A,B$ and $C$ 
for the model 1 (dashed-line) and model 2 (solid-line).
In the model 1: $C=-A$.}
\end{figure}

Once  the coefficients $A,B$, and $C$,
are determined, we can predict the amplitudes  
for the the transitions
$\gamma^\ast N \to N(1650)$, 
$\gamma^\ast N \to N(1700)$,
$\gamma^\ast N \to \Delta(1620)$ and 
$\gamma^\ast N \to \Delta(1700)$.
Based on the amplitudes used in the calibration 
we expect the estimates to be accurate for $Q^2 \gtrsim 2$ GeV$^2$~\cite{SQTM}.
%Based on the quality of the study of
%the $\gamma^\ast N \to N(1535)$, 
%$\gamma^\ast N \to N(1520)$, we expect 
%the estimates to be accurate for $Q^\gtrsim 2$ GeV$^2$~\cite{SQTM}.

From the relations (\ref{eqABC}) 
we can conclude in the limit where
no meson cloud is considered ($A_{3/2}^{D13}=0$) one has $C=-A$.
That case defines the our model 1,
and only the parameters $A$ and $B$ are necessary (since $C=-A$).
When we have a (non-zero) parametrization 
for $A_{3/2}^{D13}$, we define the model 2.
Using the parametrization discussed in 
the previous section we obtain the results 
for $A,B$ and $C$ presented in figure~\ref{figABC}.

\vspace{-.2cm}

\section{Results}

\begin{figure}
\centerline{
\mbox{%\hspace{7cm}
%\hspace{7.5cm}
\includegraphics[width=2.0in]{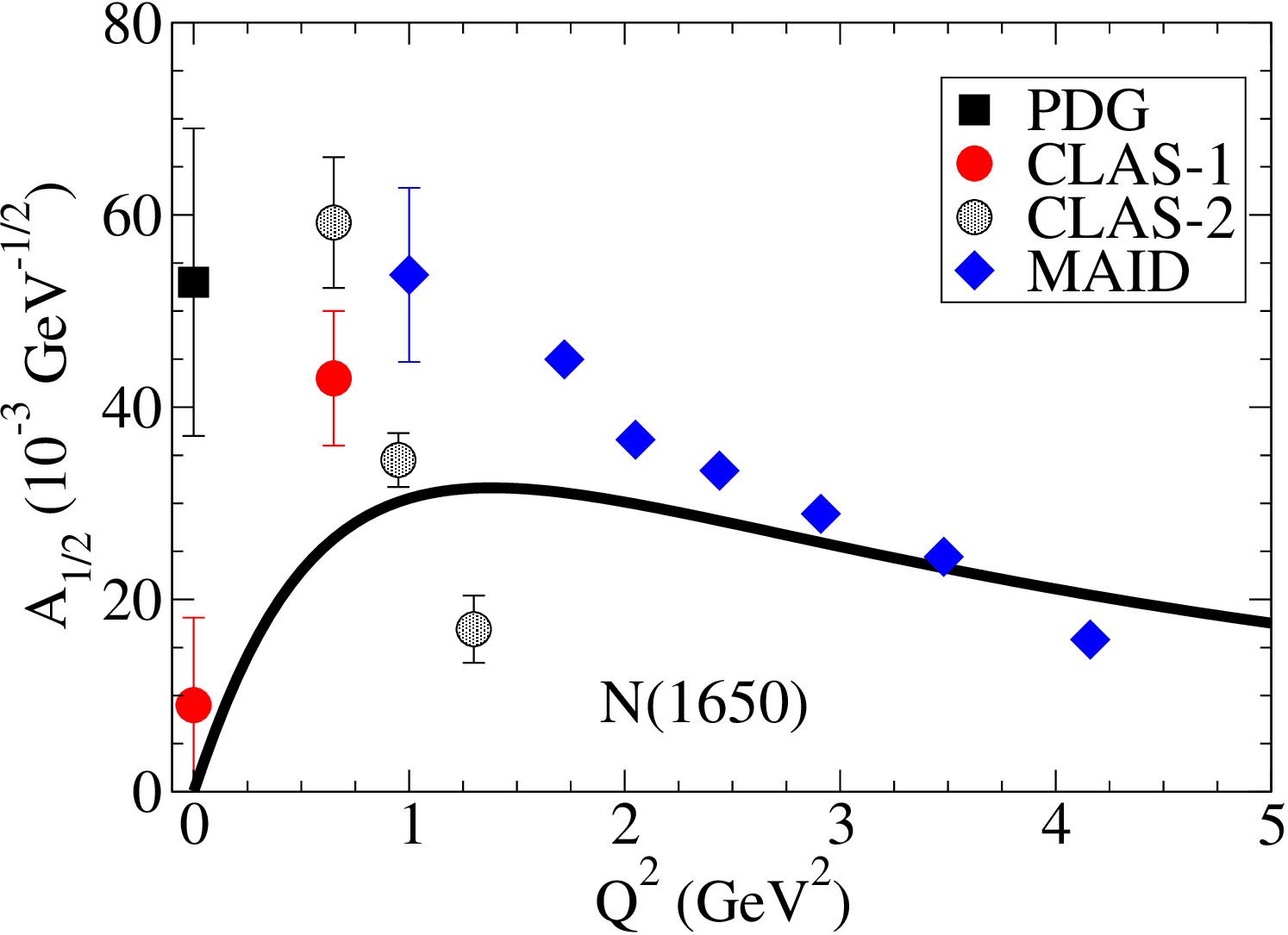}  \hspace{.1cm}
\includegraphics[width=2.0in]{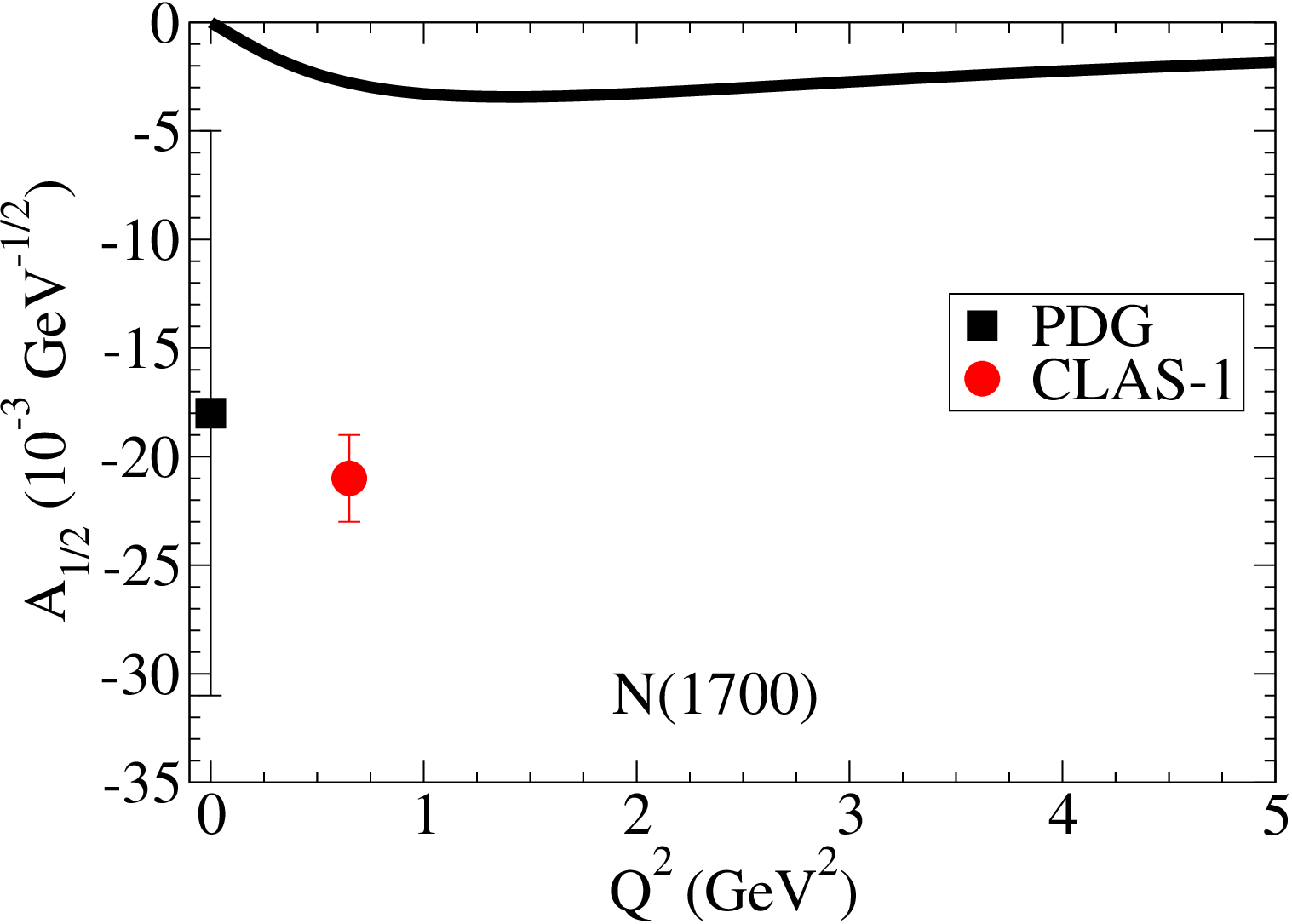}  \hspace{.1cm}
\includegraphics[width=2.0in]{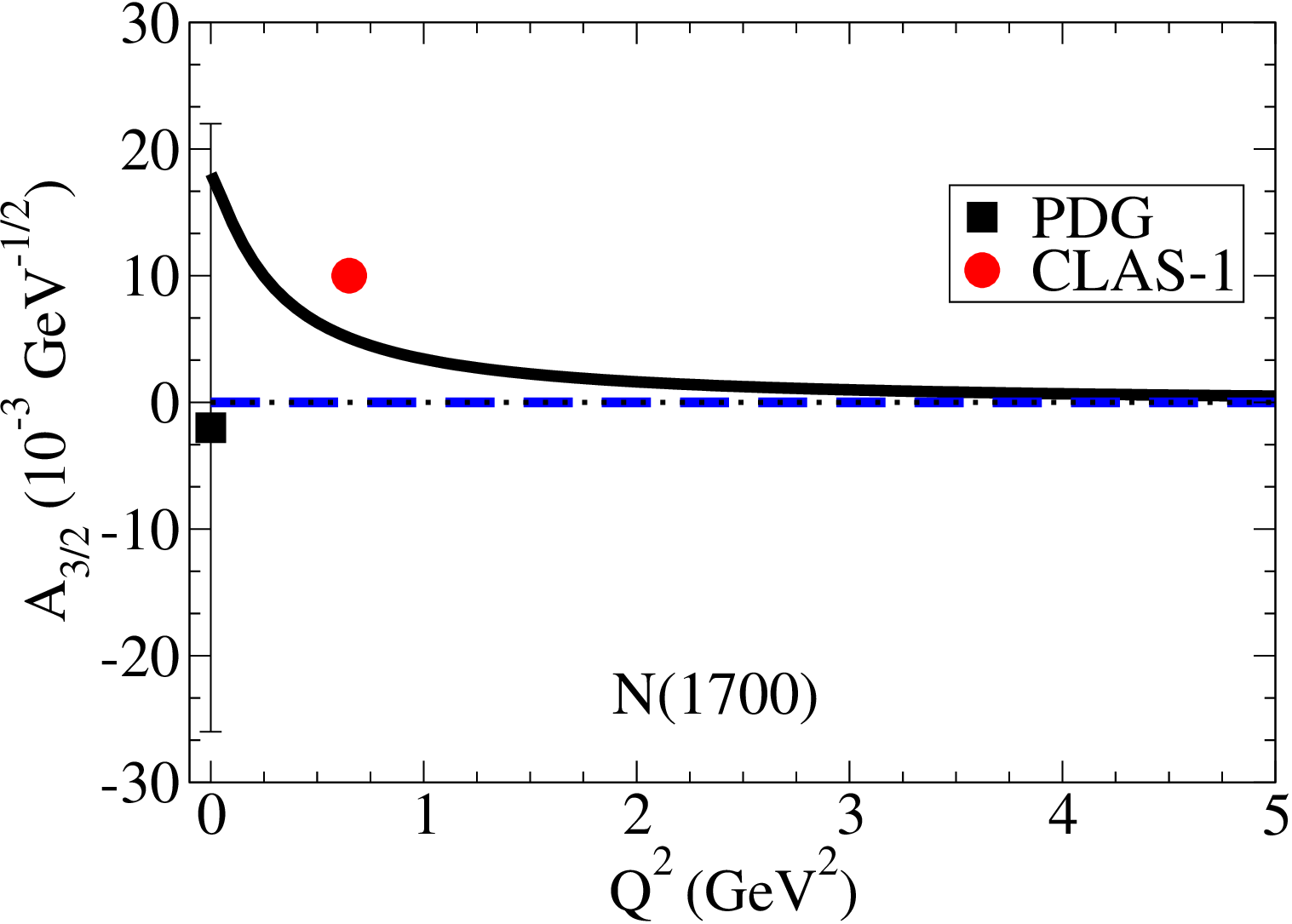} }}
\caption{\label{figN}Results for the resonances 
$N(1650)$ and $N(1700)$.
Model 1 (dashed-line) and Model 2 (solid-line).
Data from~\cite{CLAS,Dugger09,CLAS2,PDG}.}
\vspace{.3cm}
\end{figure}

\begin{figure}
%\centerline{
%\mbox{
\includegraphics[width=2.0in]{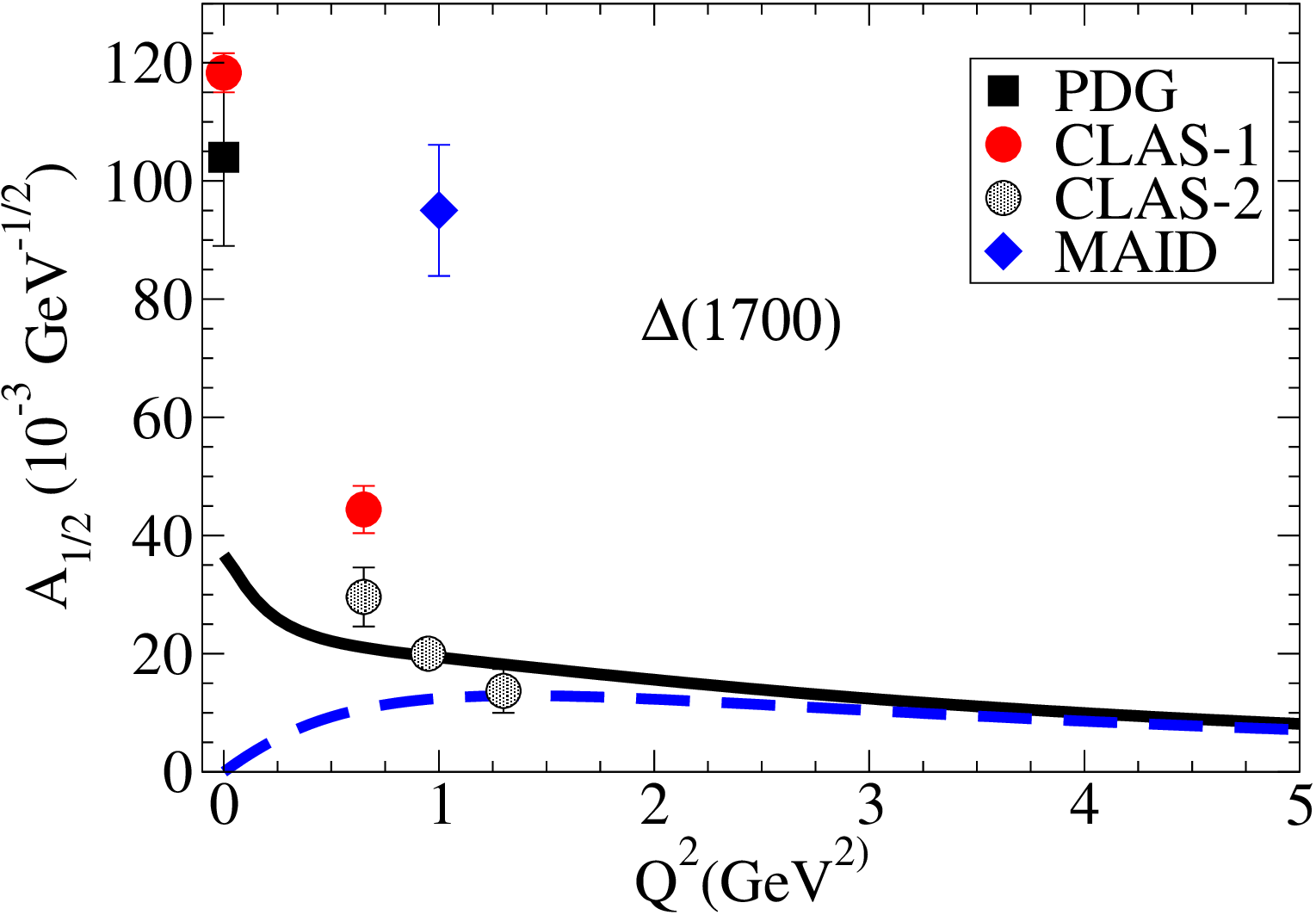}  \hspace{.1cm}
\includegraphics[width=2.0in]{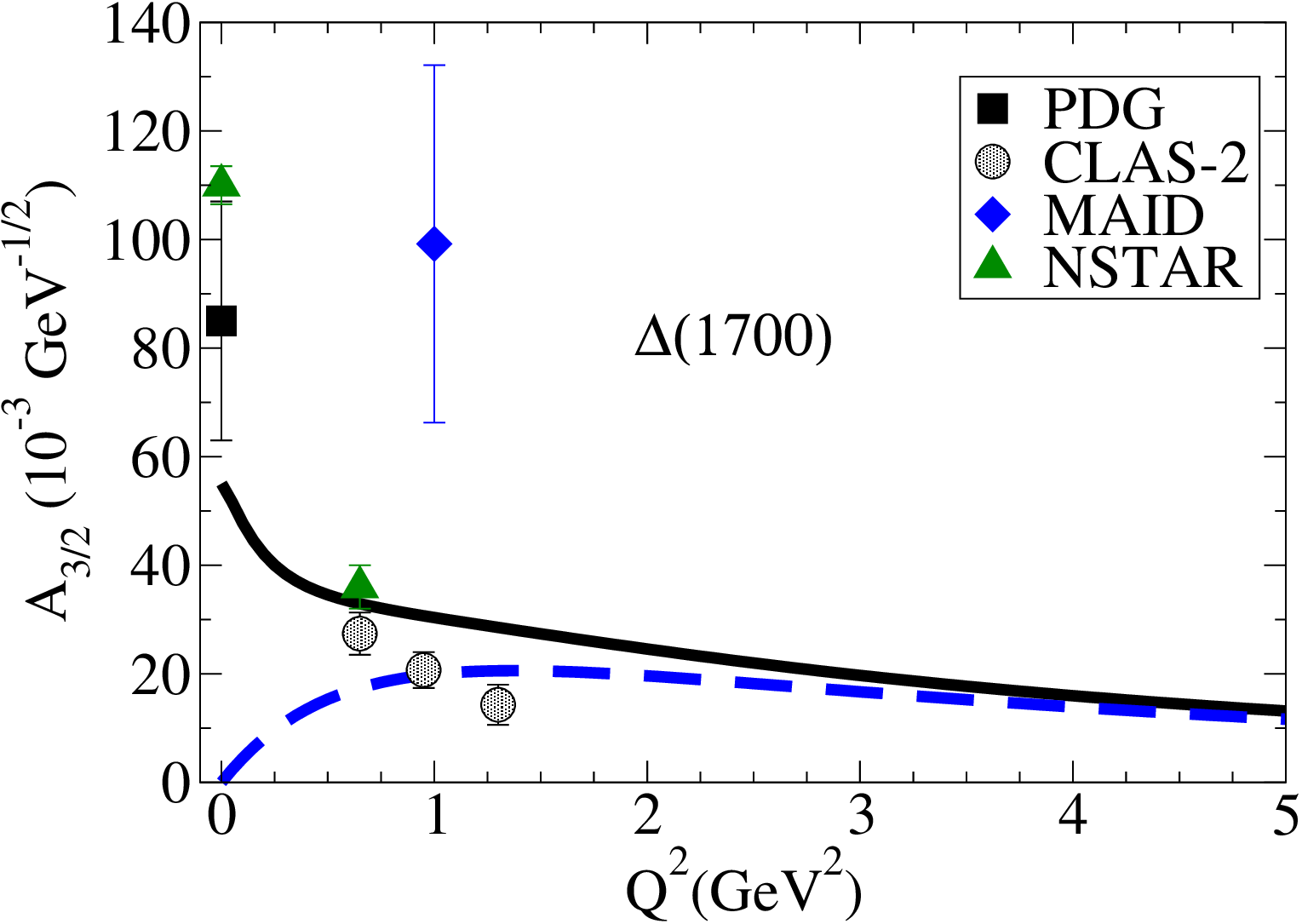} %}
\hspace{.3cm}
\begin{minipage}[b]{10pc}\caption{\label{figD1}
Results for the resonance $\Delta(1700)$.
Model 1: dashed-line; Model 2: solid line.
Data from~\cite{CLAS,Dugger09,CLAS2,PDG,NSTAR_data}.
}
%\caption{Results for the resonances $\Delta(1620)$ and $\Delta(1700)$.}
\end{minipage}
\end{figure}

With the results of the functions 
$A,B$ and $C$, represented in figure~\ref{figABC},
for the model 1 (dashed-line) and model 2 (solid-line), 
it is possible to calculate the 
amplitudes for the remaining 
transition of the multiplet $[70,1^-]$ 
using the relations from table~\ref{tableAmp}.

The results are compared with data from CLAS (CLAS-1)
\cite{CLAS,Dugger09}, preliminary data from CLAS 
(CLAS-2)~\cite{CLAS2}, data from the MAID analysis~\cite{MAID},
data presented in proceedings and  workshops (NSTAR)
\cite{NSTAR,Burkert03} and PDG data for $Q^2=0$~\cite{PDG}.

\begin{figure}[t]
%\vspace{.3cm}
\begin{minipage}{18pc}
\includegraphics[width=18pc]{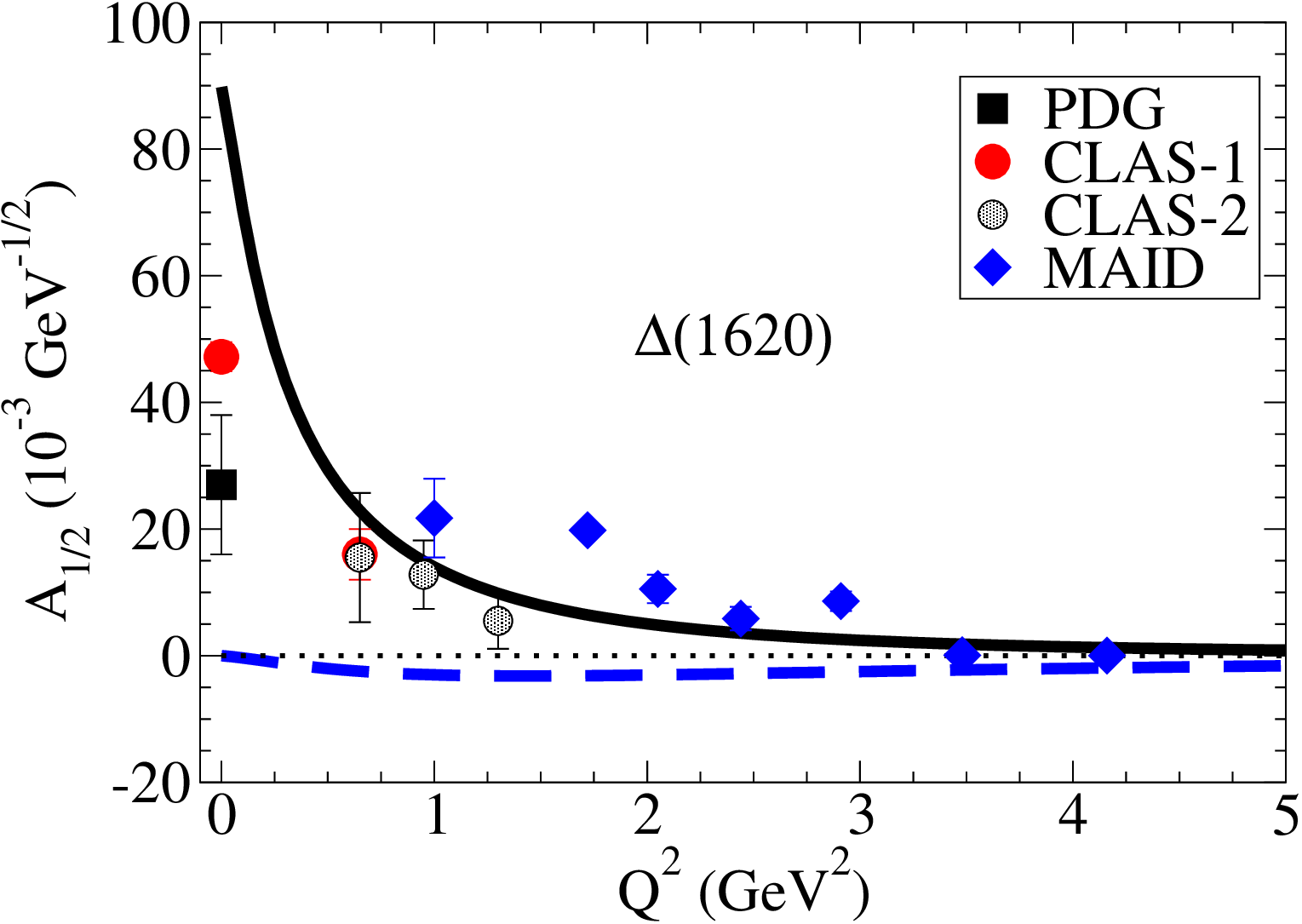}
\caption{\label{figD2}Results for $\Delta(1620)$.
Note that only the model 2 (solid-line) has 
the correct sign.}
\end{minipage}\hspace{2pc}%
\begin{minipage}{18pc}
\includegraphics[width=17pc]{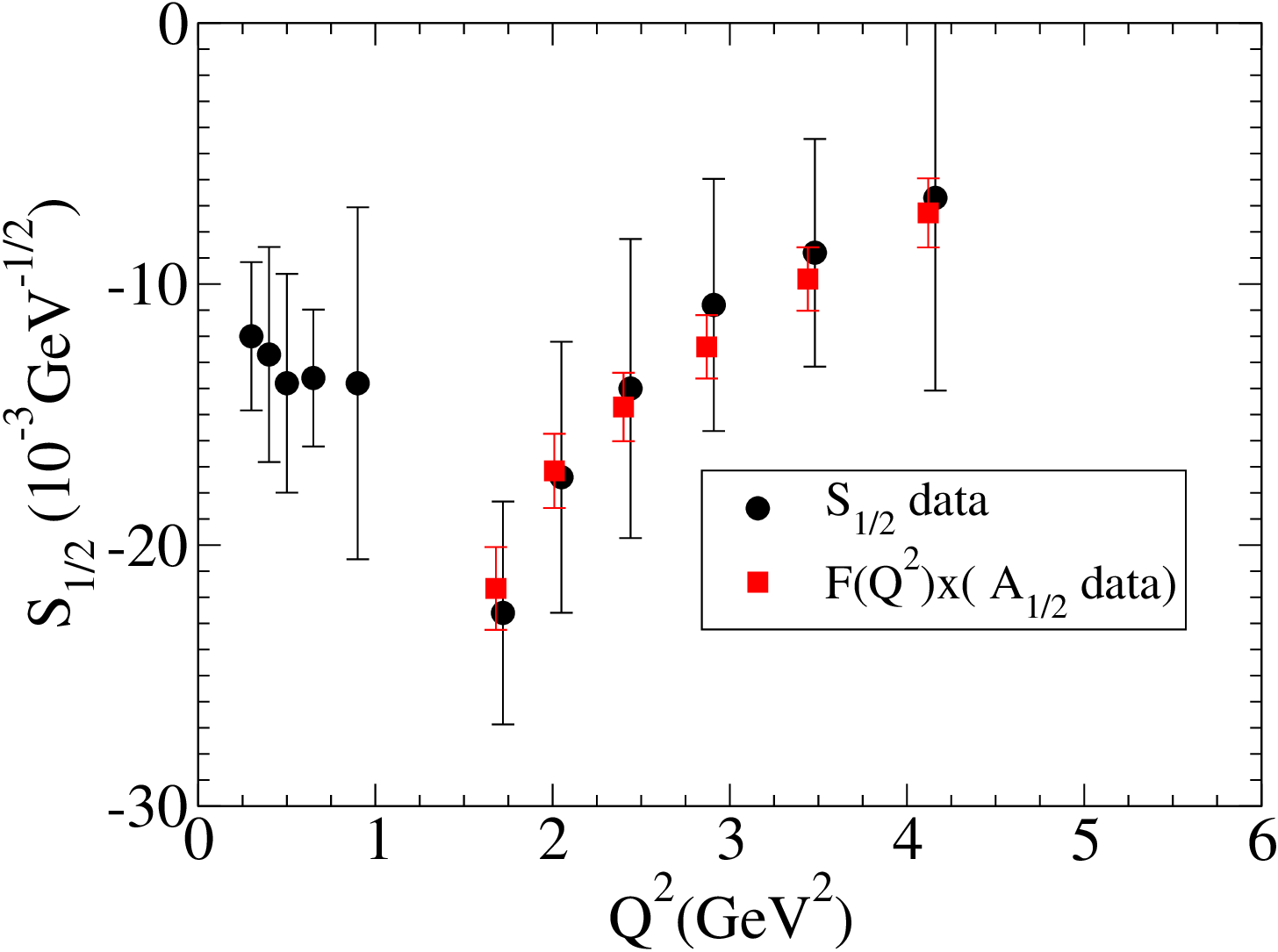}
\caption{\label{figScaling}Relation between the 
amplitudes $A_{1/2}$ and $S_{1/2}$ for 
the $\gamma^\ast N \to N(1535)$ transition. 
$F(Q^2)= - \frac{\sqrt{1 + \tau}}{\sqrt{2}} \frac{M_S^2-M^2}{2M_S Q}$.}
\end{minipage} 
\end{figure}

The results for the states $N(1650)$ and $N(1700)$
are on figure~\ref{figN}.
In the graphs, we can see that 
the model 2 gives a better description 
for the amplitude $A_{3/2}$
(the model 1 gives $A_{3/2} \equiv 0$).
Both models have the same result for 
$N(1650)$ which describe well the MAID data for $Q^2 \gtrsim  2$ GeV$^2$.
The results for the $\Delta$ states 
are presented in the figures \ref{figD1} and \ref{figD2}, respectively for 
$\Delta(1700)$ and $\Delta(1650)$. 
For $\Delta(1700)$ we can see that the 2 models have 
very similar results for $Q^2 \gtrsim 1$ GeV$^2$.
As for  $\Delta(1650)$ only the model 2 
has a good description of the data
for  $Q^2 \gtrsim 1$ GeV$^2$.
The model 1 predicts negative 
values for the amplitude $A_{1/2}$.

We can conclude then, in general, that only the model 2 
gives a good description 
of the data, particularly for $Q^2 \gtrsim 2$ GeV$^2$.
Note that the model 2 is the model that takes 
into account  the meson cloud effects 
($A_{3/2}^{D13} \ne 0$).

Based on the expected behavior 
for large $Q^2$ given by 
$A_{1/2} \propto 1/Q^3$ 
and $A_{1/2} \propto 1/Q^5$ in accordance
with perturbative QCD arguments~\cite{Carlson},
we parametrize the amplitudes as
\ba
A_{1/2}(Q^2) = D \left(\frac{\Lambda^2}{\Lambda^2 + Q^2}\right)^{3/2},
\hspace{1cm}
A_{3/2}(Q^2) = D \left(\frac{\Lambda^2}{\Lambda^2 + Q^2}\right)^{5/2},
\label{eqLargeQ2}
\ea 
for $Q^2 \approx 5$ GeV$^2$.
In the previous expression $D$ and $\Lambda$ 
are respectively coefficients and cutoffs
dependent on the amplitude and on the resonance. 
The results of the parametrizations are in table~\ref{tableLargeQ2}. 
Those results may be useful to 
compare with future experiments at large $Q^2$
as the ones predicted for the Jlab-12 GeV upgrade~\cite{NSTAR}. 

For the amplitude $A_{1/2}$ associated 
with the $\Delta(1620)$ state it was not possible 
to find a parametrization consistent the 
power $3/2$ as in equation (\ref{eqLargeQ2}).
This happens because for that 
particular amplitude there is a 
partial cancellation between the 
leading terms (on $1/Q^3$) of our $A,B$ and $C$ parametrization 
due to the difference of sign between 
the amplitudes $A_{1/2}^{S11}$ and $A_{1/2}^{D13}$
used in the determination of the SQTM coefficients 
(see dashed-line on figure \ref{figD2}).
As consequence the amplitude 
$A_{1/2}$ for the state $\Delta(1620)$ 
is dominated by next leading terms  (on $1/Q^5$)
or contributions due to meson cloud effects ($A_{3/2}^{D13}$).
It is clear in figure \ref{figD2} that, 
when we neglect the contributions from $A_{3/2}^{D13}$,
the result correspondent to the model 1 (dashed-line) is almost zero. 
This result shows that in the 
$\gamma^\ast N \to \Delta(1620)$ transition, 
contrarily to what is usually expected, 
there is a strong suppression of the 
valence quark effects for $Q^2=1$--2 GeV$^2$.
For a more detailed discussion see~\cite{SQTM}. 
A simple parametrization 
of the amplitudes $A_{1/2}$ derived 
from our model is 
$A_{1/2}= 77.21\left( \frac{\Lambda^2}{\Lambda^2 + Q^2}\right)^{5/2}$
in units $10^{-3}$ GeV$^{-1/2}$, with $\Lambda^2=1$ GeV$^2$.
Note in particular the power $5/2$, instead of the expected $3/2$.

\begin{table}[t]
\begin{tabular}{c c c c}
\hline
\hline
State & Amplitude &  $D(10^{-3}$GeV$^{-1/2}$) & $\Lambda^2$(GeV$^2$)   \\
\hline
$N(1650)$ & $A_{1/2}$ & 68.90 &  3.35\\ [.3cm]
$\Delta(1620)$ & $A_{1/2}$ &  ... & \sp ... \\ [.3cm]
$N(1700)$ & $A_{1/2}$ &  $-8.51$\sp\sp & 2.82 \\
& $A_{3/2}$ & 4.36 & 3.61   \\ [.3cm]
$\Delta(1700)$ & $A_{1/2}$ &  39.22  & 2.69 \\ 
& $A_{3/2}$ & 42.15 & 8.42   \\ [.1cm]
\hline
\hline
\end{tabular}
\hspace{.3cm}
\begin{minipage}[b]{14pc}\caption{\label{tableLargeQ2}
Parameters from the high $Q^2$ parametrization
given by equations~(\ref{eqLargeQ2}).}
\end{minipage}
\end{table}

Another interesting prediction relatively 
to the helicity amplitudes of baryons with negative parity
is the correlation between the 
amplitudes $A_{1/2}$ and $S_{1/2}$ 
associated with the $\gamma^\ast N \to N(1535)$ transition.
The consequence 
of the experimental result, $F_2^\ast \approx 0$,
observed for  $Q^2 \gtrsim 1.5$ GeV$^2$ is that  
we can write in that regime
$S_{1/2} = - \frac{\sqrt{1 + \tau}}{\sqrt{2}} \frac{M_S^2-M^2}{2M_S Q} A_{1/2}$,
where
$\tau=\frac{Q^2}{(M_S+M)^2}$~\cite{S11}.
%The correlation between the 2 amplitudes is tested in figure \ref{figScaling}.
The correlation between the 2 amplitudes 
is shown on figure \ref{figScaling}.

%Therefore those amplitudes are correlated.
%%The relation between the 2 amplitudes is tested 
%in figure \ref{figScaling}.

%\vspace{-.1cm}

\section{Summary and conclusions}
We combine the frameworks of the covariant spectator 
quark model and the single quark transition model 
in order to make predictions for 
the helicity amplitudes associated 
with negative parity resonances in the
region of masses $W=$ 1.5--1.8 GeV.
%In priciple the predictions are 
The predictions are expected to be valid 
for  $Q^2 \gtrsim 2$ GeV$^2$.
Simple parametrizations for the amplitudes are calculated
to facilitate the comparison 
with future experiments for $Q^2 \gtrsim 5$ GeV$^2$.
Contrarily to what it was expected, in some transitions, 
like for the resonances $N(1535)$ and $\Delta(1620)$, 
the valence quark effects are not dominant 
in the region $Q^2= 1$-- 2 GeV$^2$.

%\vspace{-.1cm}

\ack
The author thanks Kumar Gupta for comments and suggestions.
The author was supported by the Brazilian Ministry of Science,
Technology and Innovation (MCTI-Brazil).

%\vspace{-.1cm}

%\input{biblo2}

%\end{thereferences}


\section*{References}

\begin{thebibliography}{35}
%\begin{thereferences}


\bibitem{NSTAR} 
   Aznauryan I G, Bashir A, Braun V, Brodsky S J, Burkert V D, 
   Chang L, Chen C, El-Bennich B {\it et al.}~2013
   {\it  Int.\ J.\ Mod.\ Phys.\ E} {\bf 22}, 1330015.
%%  I.~G.~Aznauryan, A.~Bashir, V.~Braun, S.~J.~Brodsky, V.~D.~Burkert, L.~Chang, C.~Chen and B.~El-Bennich {\it et al.},
  %``Studies of Nucleon Resonance Structure in Exclusive Meson Electroproduction,''
%%  Int.\ J.\ Mod.\ Phys.\ E {\bf 22}, 1330015 (2013).
  %[arXiv:1212.4891 [nucl-th]].
  %%CITATION = ARXIV:1212.4891;%%
  %7 citations counted in INSPIRE as of 16 Jul 2013


\bibitem{Aznauryan12}
  Aznauryan I G and Burkert V D 2012  
  {\it  Prog.\ Part.\ Nucl.\ Phys.}~{\bf 67}, 1.
%%  I.~G.~Aznauryan and V.~D.~Burkert,
  %``Electroexcitation of nucleon resonances,''
%%  Prog.\ Part.\ Nucl.\ Phys.\  {\bf 67}, 1 (2012).
  %[arXiv:1109.1720 [hep-ph]].
  %%CITATION = ARXIV:1109.1720;%%
  %24 citations counted in INSPIRE as of 28 Apr 2013



\bibitem{CLAS}
  Aznauryan I G {\it et al.}  [CLAS Collaboration] 2009
  {\it Phys.\ Rev.\ C}  {\bf 80}, 055203.
%%  I.~G.~Aznauryan {\it et al.}  [CLAS Collaboration],
  %``Electroexcitation of nucleon resonances from CLAS data on single pion
  %electroproduction,''
%%  Phys.\ Rev.\  C {\bf 80}, 055203 (2009).
  %[arXiv:0909.2349 [nucl-ex]];
  %%CITATION = PHRVA,C80,055203;%%


\bibitem{MAID}
  %\nonum{MAID}
  Drechsel D, Kamalov S S, Tiator L 2007
  {\it Eur.\ Phys.\ J.\ A}  {\bf 34}, 69;
  Tiator T, Drechsel D, Kamalov S S, Vanderhaeghen M 2011
  {\it Eur.\ Phys.\ J.\ ST} {\bf 198}, 141.
%%  D.~Drechsel, S.~S.~Kamalov and L.~Tiator,
  %``Unitary Isobar Model - MAID2007,''
%%  Eur.\ Phys.\ J.\ A {\bf 34}, 69 (2007);
  %%[arXiv:0710.0306 [nucl-th]];
  %%CITATION = ARXIV:0710.0306;%%
  %% 124 citations
%%  L.~Tiator, D.~Drechsel, S.~S.~Kamalov and M.~Vanderhaeghen,
  %``Electromagnetic Excitation of Nucleon Resonances,''
%%  Eur.\ Phys.\ J.\ ST {\bf 198}, 141 (2011).
  %%[arXiv:1109.6745 [nucl-th]].
  %%CITATION = ARXIV:1109.6745;%%
  %% 9 citations 



\bibitem{Capstick00} 
  Capstick S, Roberts W 2000
  {\it Prog.\ Part.\ Nucl.\ Phys.}~{\bf 45}, S241.
%%  S.~Capstick and W.~Roberts,
  %``Quark models of baryon masses and decays,''
%%  Prog.\ Part.\ Nucl.\ Phys.\  {\bf 45}, S241 (2000).
  %%[nucl-th/0008028].
  %%CITATION = NUCL-TH/0008028;%%


\bibitem{Burkert03}
  Burkert V D, De Vita R, Battaglieri M, Ripani M 
  and Mokeev V 2003 
  {\it Phys.\ Rev.\ C} {\bf 67}, 035204.
%%  V.~D.~Burkert, R.~De Vita, M.~Battaglieri, M.~Ripani and V.~Mokeev,
  %``Single quark transition model analysis of electromagnetic nucleon resonance transitions in the [70,1-] supermultiplet,''
%%  Phys.\ Rev.\ C {\bf 67}, 035204 (2003).
  %[hep-ph/0212108].
  %%CITATION = HEP-PH/0212108;%%
  %32 citations counted in INSPIRE as of 28 Apr 2013





\bibitem{SQTM} 
  Ramalho G 2014
  {\it Phys.\ Rev.\ D} {\bf 90}, 033010.
%%  G.~Ramalho,
  %``Using the Single Quark Transition Model to predict nucleon resonance amplitudes,''
%%  Phys.\ Rev.\ D {\bf 90}, 033010 (2014).
  %[arXiv:1407.0649 [hep-ph]].
  %%CITATION = ARXIV:1407.0649;%%




\bibitem{Nucleon} 
  Gross F, Ramalho G and  Pe\~na M T 2008 
  {\it Phys.\ Rev.\ C} {\bf 77}, 015202.
%%  F.~Gross, G.~Ramalho and M.~T.~Pe\~na,
  %``A Pure S-wave covariant model for the nucleon,''
%%  Phys.\ Rev.\ C {\bf 77}, 015202 (2008).
  %[nucl-th/0606029];
  %%CITATION = NUCL-TH/0606029;%%
  %\bibitem{Nucleon2}
 




  
\bibitem{S11} 
  Ramalho G and  Pe\~na M T 2011 
  {\it Phys.\ Rev.\ D} {\bf 84}, 033007;
  Ramalho G and Tsushima K 2011
  {\it Phys.\ Rev.\ D} {\bf 84}, 051301.
%%  G.~Ramalho and M.~T.~Pe\~na,
  %``A covariant model for the gamma N -> N(1535) transition at high momentum transfer,''
%%  Phys.\ Rev.\ D {\bf 84}, 033007 (2011);
  %[arXiv:1105.2223 [hep-ph]].
  %%CITATION = ARXIV:1105.2223;%%
%%  G.~Ramalho and K.~Tsushima,
  %``A simple relation between the gamma N -> N(1535) helicity amplitudes,''
%%  Phys.\ Rev.\ D {\bf 84}, 051301 (2011).
  %[arXiv:1105.2484 [hep-ph]].
  %%CITATION = ARXIV:1105.2484;%%
  %11 citations counted in INSPIRE as of 16 Jul 2013


\bibitem{D13}
  Ramalho G and  Pe\~na M T 2014
  {\it Phys.\ Rev.\ D} {\bf 89}, 094016.
%%  G.~Ramalho and M.~T.~Pe\~na,
  %``Valence quark and meson cloud contributions to the $\gamma^\ast N \to N^\ast(1520)$ form factors,''
%%  Phys.\ Rev.\ D {\bf 89}, 094016 (2014).
  %[arXiv:1309.0730 [hep-ph]].
  %%CITATION = ARXIV:1309.0730;%%





% Spectator formalism
\bibitem{Gross}
  Gross F 1969 
  {\it Phys.\ Rev.}~{\bf 186}, 1448;
  Stadler A, Gross F and Frank M 1997
  {\it Phys.\ Rev.\  C} {\bf 56}, 2396.
%%  F.~Gross,
  %``Three-Dimensional Covariant Integral Equations For Low-Energy Systems,''
%%  Phys.\ Rev.\  {\bf 186}, 1448 (1969);
  %%CITATION = PHRVA,186,1448;
  %
  %\bibitem{Gross97}
%%  A.~Stadler, F.~Gross and M.~Frank,
  %``Covariant equations for the three-body bound state,''
%%  Phys.\ Rev.\  C {\bf 56}, 2396 (1997).
  %%[arXiv:nucl-th/9703043].
  %%CITATION = PHRVA,C56,2396;%%






\bibitem{Nucleon2}
  Gross F, Ramalho G and Pe\~na M T 2012 
  {\it Phys.\ Rev.\ D} {\bf 85}, 093005.
%%  F.~Gross, G.~Ramalho and M.~T.~Pe\~na,
  %``Covariant nucleon wave function with S, D, and P-state components,''
%%  Phys.\ Rev.\ D {\bf 85}, 093005 (2012).
  %%[arXiv:1201.6336 [hep-ph]].
  %%CITATION = ARXIV:1201.6336;%%
  %10 citations counted in INSPIRE as of 12 Jun 2013




\bibitem{OctetFF}
  Ramalho G and Tsushima K 2011
  {\it Phys.\ Rev.\ D} {\bf 84}, 054014;
  Ramalho G, Tsushima K and Thomas A W 2013
  {\it J.\ Phys.\ G} {\bf 40}, 015102.
%%  G.~Ramalho and K.~Tsushima,
  %``Octet baryon electromagnetic form factors in a relativistic quark model,''
  Phys.\ Rev.\ D {\bf 84}, 054014 (2011);
  %%[arXiv:1107.1791 [hep-ph]];
  %%CITATION = ARXIV:1107.1791;%%
%%    G.~Ramalho, K.~Tsushima and A.~W.~Thomas,
  %``Octet Baryon Electromagnetic form Factors in Nuclear Medium,''
%%  J.\ Phys.\ G {\bf 40}, 015102 (2013).
  %%[arXiv:1206.2207 [hep-ph]];
  %%CITATION = ARXIV:1206.2207;%%



\bibitem{Omega}
  Ramalho G, Tsushima K and Gross F 2009
  {\it Phys.\ Rev.\  D} {\bf 80}, 033004.
%%  G.~Ramalho, K.~Tsushima and F.~Gross,
  %``A relativistic quark model for the Omega- electromagnetic form factors,''
%%  Phys.\ Rev.\  D {\bf 80}, 033004 (2009).
  %%[arXiv:0907.1060 [hep-ph]].
  %%CITATION = PHRVA,D80,033004;%%



\bibitem{Lattice} 
  Ramalho G and Pe\~na M T 2009
  {\it J.\ Phys.\ G} {\bf 36}, 115011.
%%  G.~Ramalho and M.~T.~Pe\~na,
  %``Nucleon and gamma N ---> Delta lattice form factors in a constituent quark model,''
%%   J.\ Phys.\ G {\bf 36}, 115011 (2009).
  %%[arXiv:0812.0187 [hep-ph]].
  %%CITATION = ARXIV:0812.0187;%%
  %25 citations counted in INSPIRE as of 22 May 2014





\bibitem{Octet2Decuplet} 
  Ramalho G and Tsushima K 2013
  {\it Phys.\ Rev.\ D} {\bf 87}, 093011;
  {\it Phys.\ Rev.\ D} {\bf 88}, 053002.
%%  G.~Ramalho and K.~Tsushima,
  %``Octet to decuplet electromagnetic transition in a relativistic quark model,''
%%  Phys.\ Rev.\ D {\bf 87}, 093011 (2013);
  %%[arXiv:1302.6889 [hep-ph]];
  %%CITATION = ARXIV:1302.6889;%%
  %4 citations counted in INSPIRE as of 31 Jan 2014
  %\bibitem{DecupletDecays} 
  %%G.~Ramalho and K.~Tsushima,
  %``What is the role of the meson cloud in the $\Sigma^{*0} \to \gamma \Lambda$ and $\Sigma^\ast \to \gamma \Sigma$ decays?,''
%%  Phys.\ Rev.\ D {\bf 88}, 053002 (2013).
  %%[arXiv:1307.6840 [hep-ph]].
  %%CITATION = ARXIV:1307.6840;%%
  %2 citations counted in INSPIRE as of 07 Feb 2014



\bibitem{Roper} 
  Ramalho G and Tsushima K 2010
  {\it Phys.\ Rev.\ D} {\bf 81}, 074020;
  Ramalho G and Tsushima K 2014
  {\it Phys.\ Rev.\ D} {\bf 89}, 073010.
%%  G.~Ramalho and K.~Tsushima,
  %``Valence quark contributions for the gamma N -> P11(1440) form factors,''
%%  Phys.\ Rev.\ D {\bf 81}, 074020 (2010);
  %%[arXiv:1002.3386 [hep-ph]];
  %%CITATION = ARXIV:1002.3386;%%
  %30 citations counted in INSPIRE as of 19 Aug 2013
  %%G.~Ramalho and K.~Tsushima,
  %``$\gamma^\ast N \to N(1710)$ transition at high momentum transfer,''
%%  Phys.\ Rev.\ D {\bf 89}, 073010 (2014).
  %%[arXiv:1402.3234 [hep-ph]].
  %%CITATION = ARXIV:1402.3234;%%





\bibitem{LatticeD} 
  Ramalho G and Pe\~na M T 2009
  {\it Phys.\ Rev.\ D} {\bf 80}, 013008. 
%%  G.~Ramalho and M.~T.~Pe\~na,
  %``Valence quark contribution for the gamma N ---> Delta 
  % quadrupole transition extracted from lattice QCD,''
%%  Phys.\ Rev.\ D {\bf 80}, 013008 (2009).
  %%[arXiv:0901.4310 [hep-ph]];
  %%CITATION = ARXIV:0901.4310;%%
  %26 citations counted in INSPIRE as of 19 Aug 2013


\bibitem{Delta2} 
  Ramalho G, Pe\~na M T and Gross F 2008
  {\it Eur.\ Phys.\ J.\ A} {\bf 36}, 329;
  {\it Phys.\ Rev.\ D} {\bf 78}, 114017;
  Ramalho G, Pe\~na M T and Gross F 2010
  {\it Phys.\ Rev.\ D} {\bf 81}, 113011;
  Ramalho G, Pe\~na M T and Stadler A 2012
  {\it Phys.\ Rev.\ D} {\bf 86}, 093022.
%%  G.~Ramalho, M.~T.~Pe\~na and F.~Gross,
  %``A Covariant model for the nucleon and the Delta,''
%%  Eur.\ Phys.\ J.\ A {\bf 36}, 329 (2008);
  %%[arXiv:0803.3034 [hep-ph]];
  %%CITATION = ARXIV:0803.3034;%%
  %39 citations counted in INSPIRE as of 14 Feb 2014
  %\bibitem{NDeltaD} 
  %%%G.~Ramalho, M.~T.~Pe\~na and F.~Gross,
  %``D-state effects in the electromagnetic N Delta transition,''
%%  Phys.\ Rev.\ D {\bf 78}, 114017 (2008);
  %%[arXiv:0810.4126 [hep-ph]];
  %%CITATION = ARXIV:0810.4126;%%
  %33 citations counted in INSPIRE as of 19 Aug 2013
  %\bibitem{DeltaDFF} 
  %%G.~Ramalho, M.~T.~Pe\~na and F.~Gross,
  %``Electromagnetic form factors of the Delta with D-waves,''
%%  Phys.\ Rev.\ D {\bf 81}, 113011 (2010);
  %%[arXiv:1002.4170 [hep-ph]];
  %%CITATION = ARXIV:1002.4170;%%
  %26 citations counted in INSPIRE as of 19 Aug 2013
  %\bibitem{DeltaDeformation} 
%%  G.~Ramalho, M.~T.~Pe\~na and A.~Stadler,
  %``The shape of the $\Delta$ baryon in a covariant spectator quark model,''
%%  Phys.\ Rev.\ D {\bf 86}, 093022 (2012).
  %%[arXiv:1207.4392 [nucl-th]];
  %%CITATION = ARXIV:1207.4392;%%
  %2 citations counted in INSPIRE as of 19 Aug 2013


\bibitem{Delta1600} 
  Ramalho G and Tsushima K 2010
  {\it Phys.\ Rev.\ D} {\bf 82}, 073007.
%%  G.~Ramalho and K.~Tsushima,
  %``A Model for the $\Delta(1600)$ resonance and $\gamma N -> \Delta(1600)$ transition,''
%%  Phys.\ Rev.\ D {\bf 82}, 073007 (2010).
  %%[arXiv:1008.3822 [hep-ph]].
  %%CITATION = ARXIV:1008.3822;%%
  %17 citations counted in INSPIRE as of 19 Aug 2013





\bibitem{Strange}
  Ramalho G and Tsushima K 2012
  {\it Phys.\ Rev.\ D} {\bf 86}, 114030;
  Ramalho G and  Pe\~na M T 2011  
  {\it Phys.\ Rev.\ D} {\bf 83}, 054011;
  Ramalho G, Jido D and Tsushima K 2012
  {\it Phys.\ Rev.\ D} {\bf 85}, 093014.
  %%G.~Ramalho and K.~Tsushima,
  %``Covariant spectator quark model description of the $\gamma^\ast \Lambda \to \Sigma^0$ transition,''
%%  Phys.\ Rev.\ D {\bf 86}, 114030 (2012);
  %%[arXiv:1210.7465 [hep-ph]];
  %\bibitem{Omega2} 
%%  G.~Ramalho and M.~T.~Pe\~na,
  %``Extracting the Omega- electric quadrupole moment from lattice QCD data,''
%%  Phys.\ Rev.\ D {\bf 83}, 054011 (2011);
  %%[arXiv:1012.2168 [hep-ph]];
  %%CITATION = ARXIV:1012.2168;%%
  %11 citations counted in INSPIRE as of 07 Feb 2014
%%  G.~Ramalho, D.~Jido and K.~Tsushima,
  %``Valence quark and meson cloud contributions for the gamma* Lambda -> Lambda* and gamma* Sigma0 -> Lambda* reactions,''
%%  Phys.\ Rev.\ D {\bf 85}, 093014 (2012).
  %%[arXiv:1202.2299 [hep-ph]]
  %%CITATION = ARXIV:1202.2299;%%
  %6 citations counted in INSPIRE as of 19 Aug 2013





\bibitem{Timelike}
  Ramalho G and Pe\~na M T 2012
  {\it Phys.\ Rev.\ D} {\bf 85}, 113014.
%%  G.~Ramalho and M.~T.~Pe\~na,
  %``Timelike gamma* N -> Delta form factors and Delta Dalitz decay,''
%%  Phys.\ Rev.\ D {\bf 85}, 113014 (2012).
  %%[arXiv:1205.2575 [hep-ph]].
  %%CITATION = ARXIV:1205.2575;%%
  %9 citations counted in INSPIRE as of 19 Aug 2013





\bibitem{NucleonDIS}
  Gross F, Ramalho G, Pe\~na M T 2012
  {\it  Phys.\ Rev.\ D} {\bf 85}, 093006.
%%  F.~Gross, G.~Ramalho and M.~T.~Pe\~na,
  %``Spin and angular momentum in the nucleon,''
%%   Phys.\ Rev.\ D {\bf 85}, 093006 (2012).
  %%[arXiv:1201.6337 [hep-ph]].
  %%CITATION = ARXIV:1201.6337;%%
  %12 citations counted in INSPIRE as of 12 Jan 2015



\bibitem{OctetAxial}
 Ramalho G and Tsushima K, to be submitted.





\bibitem{EBAC}
  Sato T, Lee T-S L 2009
  {\it J.\ Phys.\ G} {\bf 36}, 073001.
%%  T.~Sato and T.~-S.~H.~Lee,
  %``Dynamical Models of the Excitations of Nucleon Resonances,''
%%  J.\ Phys.\ G {\bf 36}, 073001 (2009).
  %%[arXiv:0902.3653 [nucl-th]].
  %%CITATION = ARXIV:0902.3653;%%
  %13 citations counted in INSPIRE as of 05 Jan 2014





\bibitem{SQTM-refs}
  Hey A J G and Weyers 1974
  {\it Phys.\ Lett.\ B} {\bf 48}, 69;
  Cottingham W N and  Dunbar I H 1979
  {\it Z.\ Phys.\ C} {\bf 2}, 41.
%%  A.~J.~G.~Hey and J.~Weyers,
  %``Quarks and the helicity structure of photoproduction amplitudes,''
%%  Phys.\ Lett.\ B {\bf 48}, 69 (1974);
  %%CITATION = PHLTA,B48,69;%%
  %74 citations counted in INSPIRE as of 28 Apr 2013
%%    W.~N.~Cottingham and I.~H.~Dunbar,
  %``Baryon Multipole Moments In The Single Quark Transition Model,''
%%  Z.\ Phys.\ C {\bf 2}, 41 (1979).
  %%CITATION = ZEPYA,C2,41;%%
  %17 citations counted in INSPIRE as of 28 Apr 2013






% DATA  ---------------------------------------

\bibitem{Dugger09}  %Q2=0
  Dugger M {\it et al.}  [CLAS Collaboration] 2009
  {\it Phys.\ Rev.\ C} {\bf 79}, 065206.
%%  M.~Dugger {\it et al.}  [CLAS Collaboration],
  %``pi+ photoproduction on the proton for photon energies from 0.725 to 2.875-GeV,''
%%  Phys.\ Rev.\ C {\bf 79}, 065206 (2009).
  %%[arXiv:0903.1110 [hep-ex]].
  %%CITATION = ARXIV:0903.1110;%%
  %52 citations counted in INSPIRE as of 18 Aug 2013


\bibitem{CLAS2}
  Mokeev V I {\it et al.}  [CLAS Collaboration] 2012
  {\it Phys.\ Rev.\ C} {\bf 86}, 035203.
%%  V.~I.~Mokeev {\it et al.}  [CLAS Collaboration],
  %``Experimental Study of the $P_{11}(1440)$ and $D_{13}(1520)$ resonances from CLAS data on $ep \rightarrow e'\pi^{+} \pi^{-} p'$,''
%%  Phys.\ Rev.\ C {\bf 86}, 035203 (2012).




\bibitem{PDG} 
  Beringer J {\it et al.}  [Particle Data Group Collaboration] 2012
  {\it Phys.\ Rev.\ D} {\bf 86}, 010001.
  %J.~Beringer {\it et al.}  [Particle Data Group Collaboration],
  %``Review of Particle Physics (RPP),''
  %{\it Phys.\ Rev.\ D} {\bf 86}, 010001 (2012).
  %%CITATION = PHRVA,D86,010001;%%
  %2022 citations counted in INSPIRE as of 18 Aug 2013

\bibitem{NSTAR_data}
  Electromagnetic N-N* Transition Form
  Factors Workshop, Jlab, Newport News, 2008 (unpublished)






\bibitem{Carlson}
  Carlson C E and  Poor J L 1998
  {\it  Phys.\ Rev.\ D} {\bf 38}, 2758.
%%  C.~E.~Carlson and J.~L.~Poor,
  %``Distribution Amplitudes And Electroproduction Of The Delta And Other Low Lying Resonances,''
%%  Phys.\ Rev.\ D {\bf 38}, 2758 (1988).
  %%CITATION = PHRVA,D38,2758;%%








\end{thebibliography}
\end{document}